\documentclass[usenatbib]{mnras}
\usepackage{graphicx}
\usepackage[fleqn]{amsmath}
\usepackage[T1]{fontenc}
\usepackage{ae,aecompl}
\usepackage{amssymb,amsfonts,hyperref,color,subfig}
\usepackage[normalem]{ulem}
\hypersetup{linkcolor=red,citecolor=blue,filecolor=blue,urlcolor=magenta}
\usepackage[dvipsnames]{xcolor}

\title[Migration inside a cavity]{Revisiting migration in a disc cavity to explain the high eccentricities of warm Jupiters }

\author[Debras, Baruteau \& Donati]{Florian Debras,$^{1}$\thanks{E-mail: florian.debras@irap.omp.eu}
Cl{\'e}ment Baruteau$^{1}$, Jean-Fran\c cois Donati$^{1}$
\\
$^{1}$IRAP, Universit{\'e} de Toulouse, CNRS, UPS, Toulouse, France\\
}

\date{Accepted in MNRAS}

\pubyear{T2019}
\begin{document}
\label{firstpage}
\pagerange{\pageref{firstpage}--\pageref{lastpage}}
\maketitle

\begin{abstract}
The distribution of eccentricities of warm giant exoplanets is commonly explained through planet--planet interactions, although no physically sound argument favours the ubiquity of such interactions. No simple, generic explanation has been put forward to explain the high mean eccentricity of these planets. In this paper, we revisit a simple, plausible explanation to account for the eccentricities of warm Jupiters: migration inside a cavity in the protoplanetary disc. Such a scenario allows to excite the outer eccentric resonances, a working mechanism for higher mass planets, leading to a growth in the eccentricity while preventing other, closer resonances to damp eccentricity. We test this idea with diverse numerical simulations, which show that the eccentricity of a Jupiter-mass planet around a Sun-like star can increase up to $\sim 0.4$, a value never reached before with solely planet--disc interactions. This high eccentricity is comparable to, if not larger than, the median eccentricity of warm Saturn- to Jupiter-mass exoplanets. We also discuss the effects such a mechanism would have on exoplanet observations. This scenario could  have strong consequences on the discs lifetime and the physics of inner disc dispersal, which could be constrained by the eccentricity distribution of gas giants.
\end{abstract}


\section{Introduction}
\label{sec:introduction}
Exoplanets exhibit a much wider distribution of eccentricities and orbital obliquities than the planets of our solar system, and understanding why there is such a diversity is paramount to better constrain planetary formation and evolution. In the left panel of Figure \ref{fig:histo} we show the eccentricity of the (exo-)planets with known or estimated mass to date as a function of their orbital period and the planet-to-star mass ratio $q$ (median values). Three main groups are classically identified in this plot: (i) planets with $q \lesssim 2 \times 10^{-4}$, the majority of which being super-Earths, (ii) the hot Jupiters with $q \gtrsim 5 \times 10^{-4}$ and an orbital period inferior to $\sim 10$ days, and (iii) the warm Jupiters with $q \gtrsim 5 \times  10^{-4}$ and an orbital period greater than $\sim 30$ days. The plot clearly shows that eccentricity varies substantially from one group to another, which points to different physical mechanisms that regulate the growth and damping of eccentricity across these groups:
\begin{figure*}
\includegraphics[width=0.4\hsize]{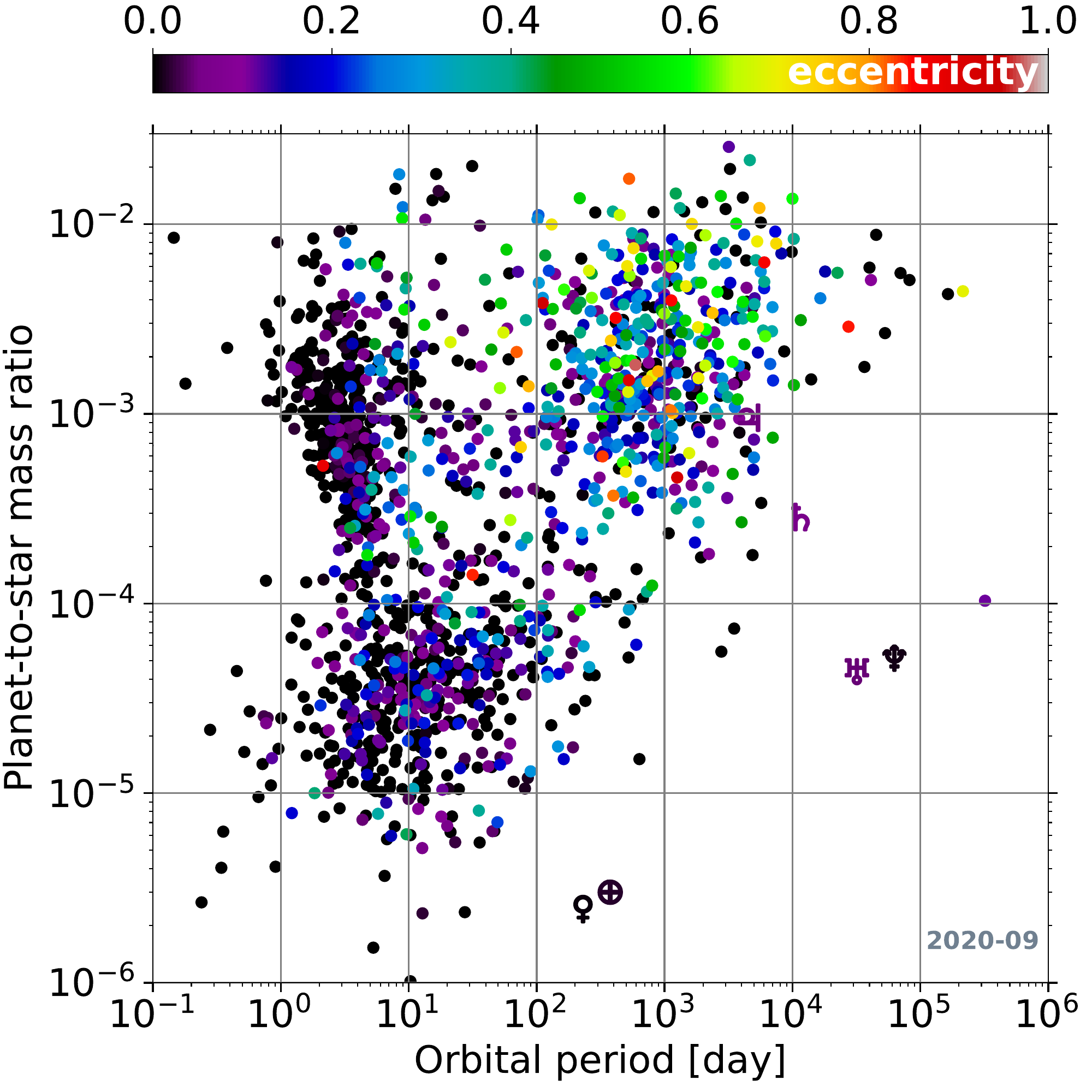}
\includegraphics[width=0.59\hsize]{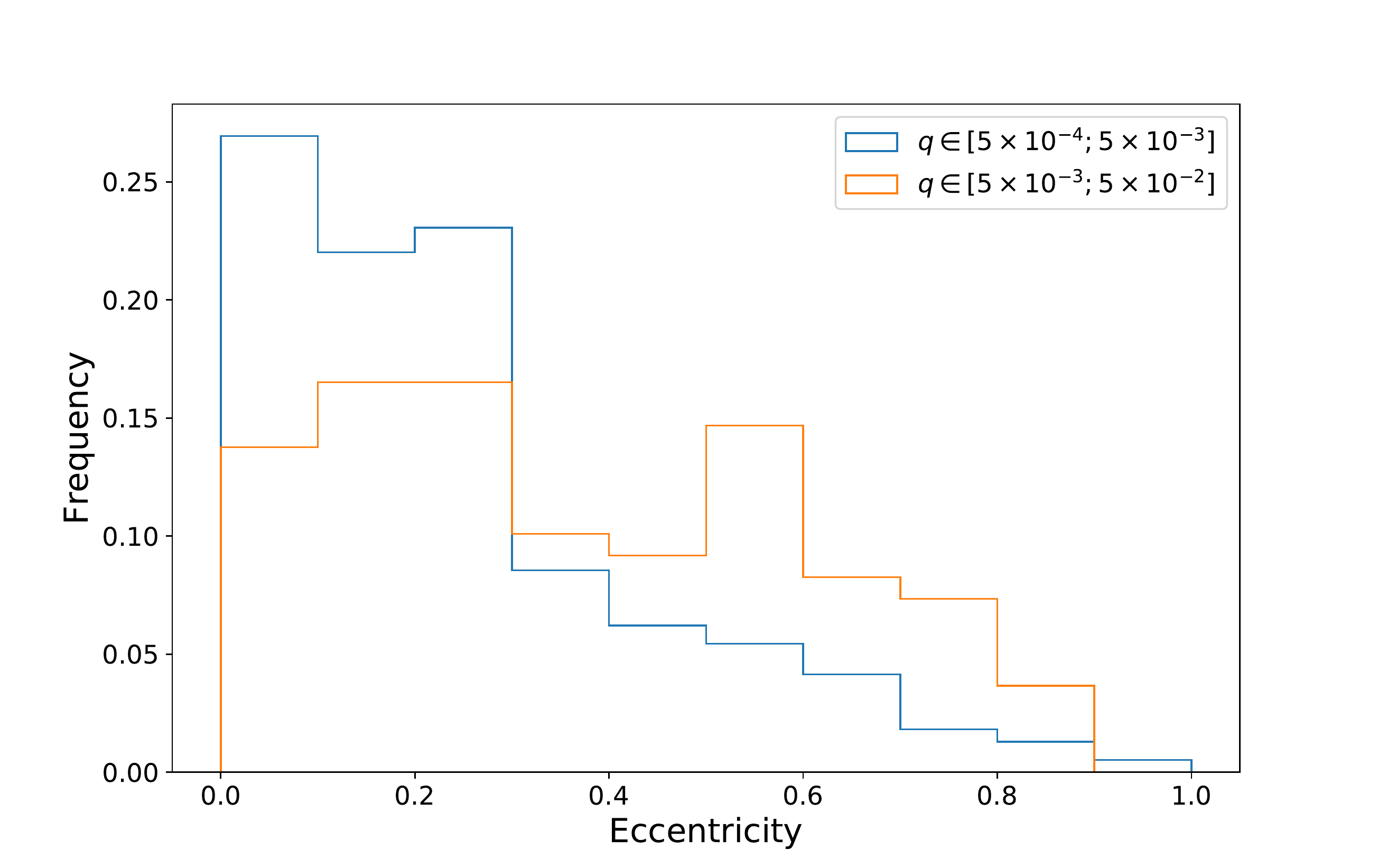}
\caption{Left: eccentricity of the exoplanets and the planets in the solar system, as a function of orbital period and planet-to-star mass ratio. Right: eccentricity distribution as a function of the planet-to-star mass ratio. All planets in this plot have an orbital period superior to 30 days to filter out planets circularized by star--planet tidal interactions. Data taken from \url{exoplanet.eu}.} 
\label{fig:histo}
\end{figure*}

\begin{enumerate}
    \item [(i)] The first group of planets where super Earths belong to typically has low eccentricities. About half of them are in multiple systems, and planet--planet interactions in (near-)resonant state 
    is the leading explanation for the low to moderate eccentricities ($e \lesssim 0.2$) in this planets population. The paradigm that the eccentricity of low-mass planets should be efficiently damped by interactions with their protoplanetary disc has been recently questioned by \citet{Fromenteau2019}, who have emphasised the need to better understand the thermal feedback from the planet on the disc (i.e. the heating of the disc due to the accretion luminosity of the planet). They have shown that there exists a critical accretion luminosity above which planets can experience an exponential increase in their eccentricity and migrate outwards, with a final value for the eccentricity of the order of the disc's aspect ratio, hence a few percent. 
    \item [(ii) ] Regarding hot Jupiters, it is thought that star--planet tidal interactions efficiently damp eccentricities and inclinations (see review in \citet{Baraffe2010}). In that regard, hot Jupiters with non-zero eccentricities are thought to be observed in the process of orbital circularisation. The question of how they acquired their initial eccentricity remains open (see e.g., \citealt{Matsumura2008, Rice2008} and \S \ref{ssec:hotJ}). 
    \item[(iii)] Finally, the third group of planets, which we 
    refer to as the warm Jupiters (Jupiter and Saturn lie at the outer, cooler,  edge of this subsample of planets, Figure \ref{fig:histo}), exhibits much higher eccentricities. The eccentricity distribution seems to be spread over the whole sample, without clear dependence on the orbital period. The right panel of Figure \ref{fig:histo} shows the eccentricity distribution of warm Jupiters as a function of $q$ (the planet-to-star mass ratio). 
    For $5 \times 10^{-4} < q < 5 \times 10^{-3}$, that is roughly between the mass of Saturn and 5 times the mass of Jupiter, the distribution of eccentricities is almost uniform up to $e \sim 0.3$, then drops and decreases slowly with increasing eccentricity. The mean eccentricity is 0.25, the median is 0.2 and $50 \%$ of the planets have $0.1 \lesssim e \lesssim 0.4$. For higher $q$ values, the  eccentricity distribution is almost uniform up to $e \sim 0.8 $ with few planets with  higher eccentricities. The mean eccentricity of this subgroup is $0.36$ while the median is 0.34
\end{enumerate}

 \citet{Papaloizou2001} showed that, for planets with $q \gtrsim 10^{-2}$, eccentricities up to $\sim$0.25 could be attained by planet--disc interactions. Other mechanisms, like planet--planet scattering or Kozai cycles, are however needed to explain larger eccentricities. \citet{Papaloizou2001} showed that planets on fixed circular orbits with masses larger than $10 M_\mathrm{J}$ (where $M_\mathrm{J}$ denotes the mass of Jupiter) carve wide enough gaps in the disc around their radial position to incorporate the principal Lindblad resonances and first order corotation resonances inside the gap (their radial location is in the range $[0.63-1.58]r_\mathrm{p}$, with $r_\mathrm{p}$ the radial position of the planet). These resonances therefore have a small effect on the torque exerted by the disc on the planet, and the evolution of the planet's orbital parameters is mostly dictated by the next, most important resonances: the first-order Lindblad resonances. The latter may lead to an exponential growth of the planet's eccentricity, controlled by the excitation of new eccentric resonances that eventually stops this eccentric instability.  

On the other hand, the eccentricity distribution of warm Jupiters with $5 \times 10^{-4}  \lesssim q \lesssim 5 \times 10^{-3} $ remains to be explained. Planet--planet scattering is often invoked \citep[see e.g.,][]{Anderson2019}. However, as stated by \citet{Armitage2011}, "there is no straightforward independent argument that the [...] initial conditions needed for [planet--planet scattering] to work are generically realized in nature", and it would probably tend to increase similarly the eccentricity of lower mass planets (which is not seen on Figure \ref{fig:histo}). Therefore, planet--planet scattering appears to be a circumstantial explanation rather than a fundamental, general situation. 

Several works have examined whether disc--planet interactions could resolve this issue for Jupiter-mass planets. Analytically, \citet{Goldreich2003} have shown that the eccentricity damping due to corotation resonances slightly exceeds the excitation from Lindblad resonances, which prevents eccentricity growth unless the corotation torque is saturated enough, and concluded that numerical simulations could help constraining the exact limit of their mecanism. Unfortunately,  when incorporating a Jupiter-mass planet in a disc with various setups (see e.g., \citealp{Bitsch2013}), the overall conclusion is that planet--disc interactions tend to damp the eccentricity $e$ to small values ($e \lesssim 0.03$). There are however two exceptions: first, when the disk is light enough, there is a tendency for eccentricity to grow up to $\sim 0.1$ over $10
^5$ orbits \citep{Ragusa2018} from secular interactions between the planet and the disc.  Secondly, when/if the planet leads to a strong decrease in the density inside its orbit, thereby carving a kind of cavity (e.g., \citealp{Dunhill2013}), the expression of eccentric resonances becomes dominant just like in \citet{Papaloizou2001}. \citet{Dangelo2006} have obtained the highest eccentricities for Jupiter-mass planets in such configuration, with $e$ reaching up to $\sim0.15$. 
\citet{Rice2008} carried out a comparable, although simpler  numerical exploration for hot Jupiters, where the density of the gas in the magnetospheric cavity was set to zero, and obtained an eccentricity increase up to $0.15$. \citet{Dangelo2006} realised that the dominant eccentric resonances are the 2:4 or the 3:5, rather than the 1:3 resonance originally proposed by \citet{Papaloizou2001}, as confirmed analytically by \citet{Teyssandier2016,Teyssandier2017}, but the question of how to obtain larger eccentricities for the lower mass range of warm Jupiters still remains open.

\citet{Duffell2015} have evaluated numerically the torque and power felt by a planet on a fixed eccentric orbit. Their conclusion is that Jupiter-mass planets can experience an increase in their eccentricity from disc--planet interactions, but this increase is limited to $e \sim 0.07$ because of interaction with non-eccentric gas. They show that, if eccentric motions in the disc are not favoured unlike in \citet{Papaloizou2001}, the planet experiences a decrease of its eccentricity from the gravitational interaction with the circular gas. A mechanism that would significantly increase the eccentricity of a warm Jupiter through planet--disc interactions therefore needs to overcome this effect. 

Motivated by this realization, in this paper we propose a plausible explanation for the eccentricity distribution of warm Jupiters: migration inside a cavity. If a cavity already exists in the inner parts of the disc, and if the planet eventually migrates inside this cavity, the exponential increase of the planet's eccentricity proposed by \citet{Papaloizou2001} could set in and the gas in the cavity would not oppose this effect because its density  is too low. 

 Our proposition is also motivated by disc observations, as discs with cavities - so called transition discs - have been extensively observed (see e.g., 
reviews in \citet{Gorti2016,Ercolano2017-r}). Notably, \citet{Carmona2017} has shown evidence that the inner 6 AU of the disc HD 139614 is depleted in  gas, and \citet{Muley2019} proposed that the shape of the PDS70 disc could be due to an eccentric super-Jupiter inside the cavity (see also \citealp{Bae2019}). Very recently, \citet{Calcino2020} also proposed that the near infrared scattered light observations of the MWC 758 disc could be explained by an eccentric, $10 M_\mathrm{J}$ planet in a cavity. However, in both cases whether the eccentric planet carved the cavity or whether the presence of the cavity excited the eccentricity of the planet is not known. 

The paper is organized as follows: in Section~\ref{sec:description}, we detail the physical model and numerical setup we have used to test the effects of migration inside a cavity. We then discuss the results and the effects of different parameters on the growth of eccentricity in Section \ref{sec:results}. Finally, in Section \ref{sec:discussion}, we discuss the implications of our results on exoplanet observations and for the evolution of protoplanetary discs. Globally, we do not aim at proposing the most realistic simulations of planet migration in a cavity, but rather at advocating that eccentricity growth is a strong, generic feature for such planets, simple and robust in the numerical sense. Eventually we obtain that migration into cavities is a powerful mechanism to pump the eccentricities of giant planets up to the observed mean eccentricity of warm Jupiters, without the need for planet--planet scattering.

\section{Physical model and numerical methods}
\label{sec:description}

\subsection{Disc model}
\label{ssec:disc}
Throughout this paper, we simulate the evolution of a planet embedded in its protoplanetary disc with the use of the FARGO3D code \citep{FARGO3D}. We have also run several simulations with the original 2D FARGO code \citep{Masset2000} for comparison, and the results were consistent in all cases. We consider a two-dimensional disc model described by polar cylindrical coordinates $\{r,\varphi\}$ in a reference frame centred on the star. The star remains fixed in the simulations, and the indirect terms due to the star motion arising from the gravity of the disc and the planet are accounted for in the evolution of the disc and of the planet. Turbulent transport of angular momentum is modelled by the use of a kinematic viscosity $\nu$, that depends on radius (see \S \ref{ssec:cavity}).

Our disc model features an inner cavity in the gas that forms independently of the planet (see Sections~\ref{ssec:cavity} and \ref{ssec:planet}). For simplicity we use a locally isothermal equation of state in our different sets of simulations, meaning that the gas temperature does not evolve with time but depends on $r$. The aspect ratio of the disc, $h = c_s/v_\mathrm{K}$, with $c_s$ the sound speed (proportional to the square root of the temperature $T$) and $v_\mathrm{K}$ the Keplerian velocity, is fixed to a constant value. The radial dependence of the temperature directly translates into a radial dependency in the aspect ratio. A uniform temperature implies $h \propto \sqrt{r}$, while a uniform aspect ratio implies $T \propto 1/r$. In our {\it 'Reference'} run (see Section \ref{sec:results} and Table \ref{tab:data}), $h$ is chosen uniform at a value of $0.05$. 

In our {\it 'Reference'} simulation, the Toomre parameter $Q = c_\mathrm{s} \Omega/(\pi \mathcal{G} \Sigma)$ -  with $\Omega$ the keplerian frequency,  $\mathcal{G}$ the gravitational constant and $\Sigma$ the surface density of the gas - is $\sim 4 \times 10^5$ in the innermost part of the cavity where the density is very low, $15$ at the outer edge of the cavity and $4$ at the outer boundary condition. Although this latest value is quite low, it relies on our assumption of a uniform, high density  outside the cavity. This allows to decrease the computational time by increasing the speed of the inward migration of the planet, but is at the limit of gravitational stability for the disc. We therefore neglect the gas self-gravity in order to capture only the physical effects of the migration while keeping a reasonable computational time. 



All quantities reported hereafter are expressed in the following code units: $M_\star$, the mass of the star, as unit of mass, and $R_\mathrm{1}$, a characteristic radial size of the cavity, as unit of length. The gravitational constant $\mathcal{G}$ is set to 1. In the results presented in the next sections, time is expressed in orbits, an orbit being defined as the orbital period at a radial distance $R_\mathrm{1}$.

\subsection{Opening and preserving a cavity}
\label{ssec:cavity}

At the beginning of the simulations, a cavity is carved through a smooth jump in the initial gas surface density profile. The initial density profile increases from $\Sigma_{\rm c} = 10^{-6}$ in the cavity to $\Sigma_0 = 10^{-3}$   in our {\it 'Reference'} simulation with a transition region at $r=r_{\rm c}$ of radial width equal to two to four times the local pressure scale height. Apart from this transition region, the initial gas surface density is therefore uniform inside and outside the cavity. 

In order to preserve a cavity during the simulation, a smooth jump in the gas kinematic viscosity is adopted. In practice, $\nu$ decreases by two orders of magnitude across $r=r_{\rm c}$. The transition is either linear in the logarithm of radius or exponential : 

\begin{equation}
\nu = \left\{
    \begin{array}{ll}
        \nu_{\rm c} & r < r_\mathrm{c} \\
        \nu_{\rm c} + (\nu_0-\nu_{\rm c})\frac{\log(r/r_{\rm c})}{\log((r_{\rm c}+ \Delta r)/r_{\rm c})}  & r_\mathrm{c} < r < r_\mathrm{c} + \Delta r\\
        &  \text{or} \\
        \nu_0\exp{(\frac{(r_{\rm c}+ \Delta r-r)}{\Delta r} \log\frac{\nu_{\rm c}}{\nu_0})}  & r_\mathrm{c} < r < r_\mathrm{c} + \Delta r\\
        \nu_{0} & r > r_\mathrm{c} + \Delta r
    \end{array}
\right.
\label{eq:visc}
\end{equation}
The use of a viscosity jump to maintain the cavity is mostly for numerical convenience, as we do not specify a mechanism for building up the cavity. Still, we note that as the density inside the cavity is much lower, we might expect its ionisation degree to be much higher and magnetic turbulence will prevail in this region, thereby contributing to a higher effective turbulent viscosity \citep{Balbus1991,Balbus1998}. On the opposite, the outer disc is more likely to be in a poorly ionised state, whereby the level of turbulence should be much lower \citep[e.g.,][]{Simon2018}. 

For our {\it 'Reference'} run, we have chosen the same value than \citet{Papaloizou2001} for the outer viscosity (equivalent to $\alpha \sim 10^{-3}$, where $\alpha$ is the classical Shakura-Sunyaev parameter \citep{Shakura1973}), which leads to $\alpha \sim 0.1$ within the cavity. \citet{Papaloizou2001} concluded that, if the viscosity in their simulations was decreased, their mechanism for eccentricity growth could be applied to lower-mass planets. Therefore, with the chosen viscosity distribution, we are not mimicking \citet{Papaloizou2001}'s work but really assessing the impact of the cavity on the orbital evolution of giant planets of Jupiter mass.


\begin{figure}
\centering
  \includegraphics[width=\linewidth]{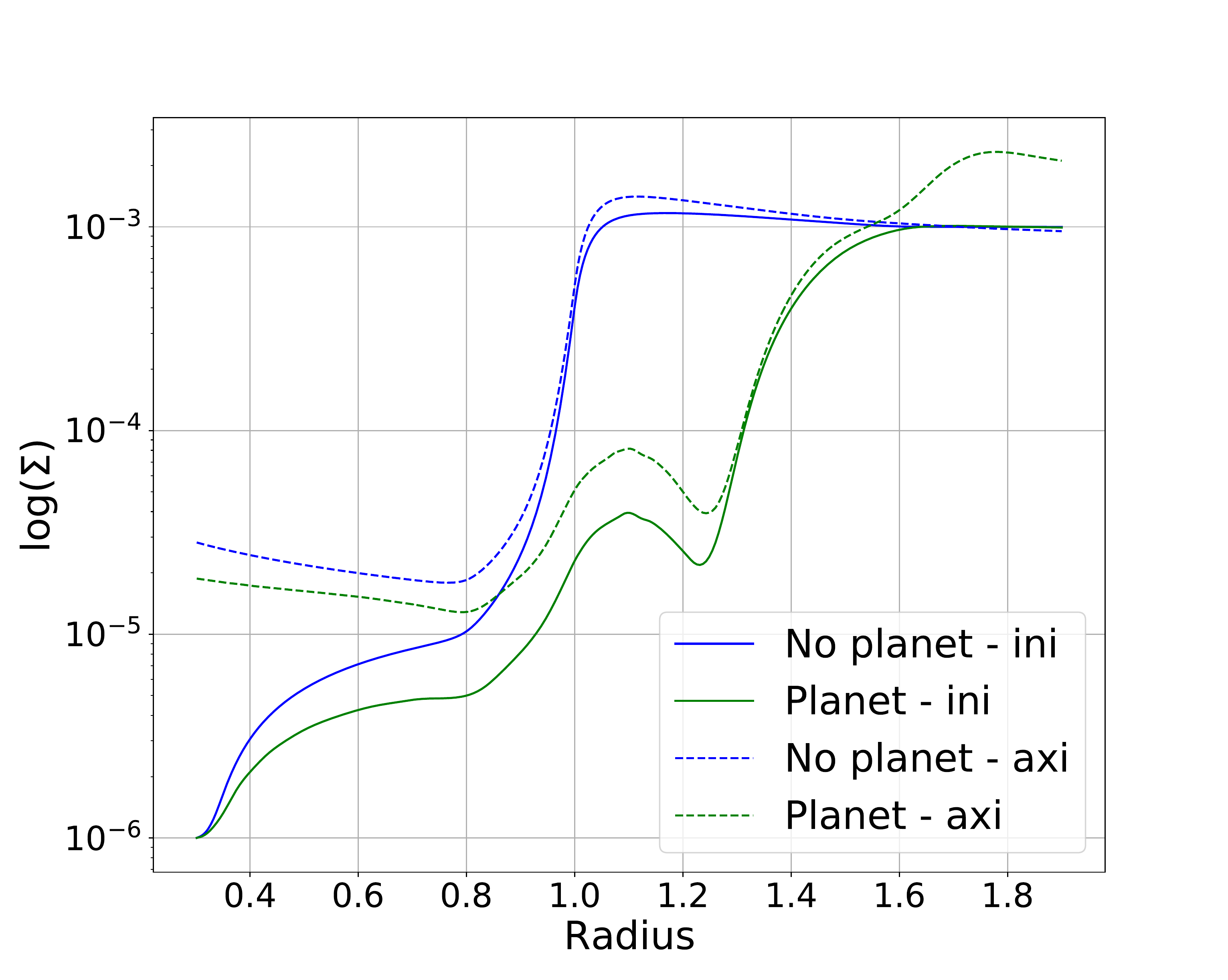}
\caption{Azimuthally averaged surface density (decimal log) as a function of radius before a planet is introduced ('No planet', blue curves), and just before we allow the planet to feel the disc, once a gap is carved ('Planet', green curves). 'Ini' or 'Axi' refer to different boundary conditions in the inner and outer wave killing zone, see Section \ref{ssec:boundary}. }
\label{fig:init_cavity}
\end{figure}


\subsection{Planet}
\label{ssec:planet}

We simulate the migration of a giant planet in the disc cavity in three steps: first, we impose an initial  axisymmetric gas density exhibiting a cavity, and allow for the disc to evolve in 1D without the presence of a planet for a thousand orbits.\footnote{This is less than the viscous timescale in the disc ($\sim 10^5$ orbits) but we have verified with a dedicated 1D simulation that was run for $3 \times 10^5$ orbits that the planetary evolution shows only marginal differences whether this first step lasts 1000 or 300 000 orbits.} The blue curves in Figure~\ref{fig:init_cavity} show the gas density profile after this first step for two choices of boundary conditions, described below. We then use this 1D profile for density and gas velocity as an initial axisymmetric profile of our 2D simulations and add a planet on a fixed circular orbit outside the cavity. The mass of the planet is gradually increased over 15 planet orbits to avoid spurious effects of incorporating brutally the planet inside the disc. This second step lasts for 600 orbits, giving enough time for the planet to carve its gap. When the planet and the cavity do not evolve significantly anymore (green lines, Figure \ref{fig:init_cavity}), we allow for the planet to feel the gas torque and migrate. In our {\it 'Reference'} run, the planet is initialised at $r_{p,\mathrm{ini}} = 1.1$. We have assessed with other simulations that the initial position does not impact the final results as long as the planet is initially outside the cavity

The gravitational potential of the planet, $\Phi_p$ is smoothed over a softening length $\epsilon = 0.5 H(r_p)$, hence:
\begin{equation}
\Phi_p = - \dfrac{G M_p}{(|\vec{r}-\vec{r_p}|^2+ \epsilon^2)^{1/2}},
\end{equation}
where $M_p$ is the mass of the planet, and $\vec{r_p}$ its position vector. Our simulations assuming a Sun-mass star, the planet-to-star mass ratio in the {\it 'Reference'} run amounts to $q=10^{-3}$, which corresponds  to a Jupiter mass planet (different values of $q$ will be considered in Section~\ref{ssec:physics}). Because the planet needs to migrate over about half its initial radial separation before its principal outer Lindblad resonances fall in the cavity, and because the migration rate decreases upon reaching the cavity, the simulations demand to be run for $10^4$ orbits at minimum. Therefore, as we want to explore the parameter space to assess the robustness of our mechanism, our {\it 'Reference'} simulation has a medium  grid resolution ($400 \times 400$ points in the radial and azimuthal directions) but most of our simulations have a low grid resolution ($200 \times 200$). Both these resolutions are not enough to properly consider the effects of the gas located within a Hill sphere on the planet. The torque exerted on the planet by the gas located within its Hill sphere is therefore neglected, which is also recommended when self-gravity of the gas is discarded \citep{Crida2009}.


\subsection{Boundary conditions and wave-killing zones}
\label{ssec:boundary}

In order to prevent spurious reflections of the planet wakes at the grid's radial edges, we impose an inner and an outer wave-killing zones. In these wave-killing zones, the density and the radial and azimuthal velocities are relaxed either to their initial values, or to their instantaneous axisymmetric values. None of these choices are perfectly representative of a physical reality, but changing from one to the other allows to assess the impact of the boundary conditions. We will refer to as 'ini' and 'axi' the damping towards initial values or axisymmetric values of the fields, respectively, as in Figure~\ref{fig:init_cavity}.

The radial size of the wave-killing zones is chosen such that the ratio of the orbital period at the boundary and at the edge of the wave-killing zones is 1.3, as in \citet{Benitez2016}. Namely, for a grid expanding from $r=0.3$ to $r=1.9$, the outer edge of the inner wave-killing zone is $r=0.36$, while the inner edge of the outer wave-killing zone is $1.6$. The damping procedure makes use of a parabolic ramp following \citet{Devalborro2006}, their Eq.(10), and the characteristic damping timescale is chosen to be 3 local orbital periods.

In addition to the use of wave-killing zones, the boundary conditions are dealt with three cells of ghost zones at both radial edges of the grid. At the inner edge of the simulation, the boundary conditions for the density and the azimuthal velocity require that ghost cells follow power-law extrapolations of the inner values. The radial velocity is an outflow boundary condition, which allows mass to flow symmetrically inside the ghost cells but the velocity is set to zero if it is directed from the ghost cells to the simulation domain. At the outer edge, power-law extrapolations are also used for the density and the azimuthal velocity, while we impose an anti-symmetric boundary condition on the radial velocity.

\section{Numerical results}
\label{sec:results}

We present in this section the results of our different sets of simulations. Table~\ref{tab:data} gathers the values of the physical parameters in the {\it 'Reference'} simulation:
\begin{enumerate}
\item[-] $N_r$ and $N_\phi$ are the number of grid cells in the radial and azimuthal directions, 
\item[-] $r_\mathrm{in}$ and $r_\mathrm{out}$ are the inner and outer radial boundaries of the simulated domain, 
\item[-] $h$ is the disc's aspect ratio, with a flaring index $f$ (namely, $h(r) \propto r^f$), 
\item[-] $r_\mathrm{c}$ denotes the characteristic width of the cavity, which extends from $r_\mathrm{in}$ to $r_\mathrm{c} + \Delta r$, with $\Delta r$ the size of the transition
region between the cavity and the outer disc (see Section \ref{ssec:cavity}), 
\item[-] $\Sigma_0$ and $\Sigma_c$ denote the initial surface densities in the outer disc and inside the cavity, respectively,
\item[-] $\nu_0$ and $\nu_c$ denote the constant kinematic viscosities in the outer disc and inside the cavity,
\item[-] the "Damping" parameter refers to the damping procedure in the wave-killing zones, see Section \ref{ssec:boundary}.
\item[-] the "visc" parameter refers to the transition of viscosity, see Eq.\eqref{eq:visc}.
\end{enumerate}

\begin{table}
\centering
\begin{tabular}{|l|l|}
  \hline
   & Reference \\
  \hline
  $N_r$ & 400 \\
  $N_\phi$ & 400 \\
  $r_\mathrm{in}$ & 0.3  \\
  $r_\mathrm{out}$ & 1.9  \\
  $h$ & 0.05 \\
  $f$ & 0\\
  $r_c$ & 0.8\\
  $\Delta r$ & 0.2 \\
  $\Sigma_0$ & $10^{-3}$ \\
  $\Sigma_c$ & $10^{-6}$\\
  $\nu_0$ & $2.5 \times 10^{-6}$\\
  $\nu_c$ & $2.5 \times 10^{-4}$\\
  Damping & ini\\ 
  visc & Linear in log \\
  \hline
\end{tabular}
\caption{Parameters used in the {\it 'Reference'} simulation (see the beginning of Section~\ref{sec:results} for the meaning of each parameter). A '0' subscript is used to denote a quantity for the gas outside the cavity, and a 'c' subscript for quantities inside the cavity.}

\label{tab:data}
\end{table}

\subsection{Reference  run}

\begin{figure*}
\centering
  \includegraphics[width=.45\linewidth]{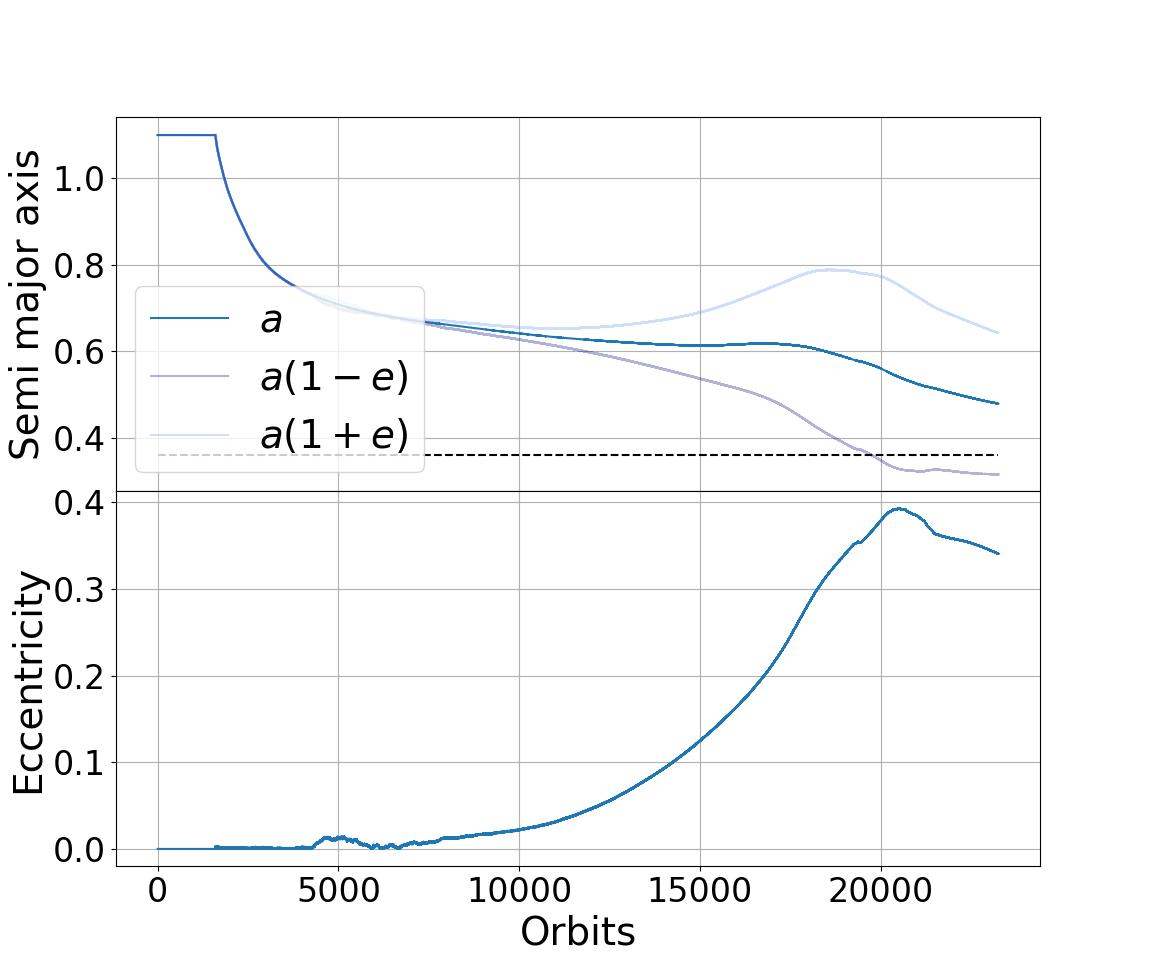}\label{fig:ref_1}
  \includegraphics[width=.45\linewidth]{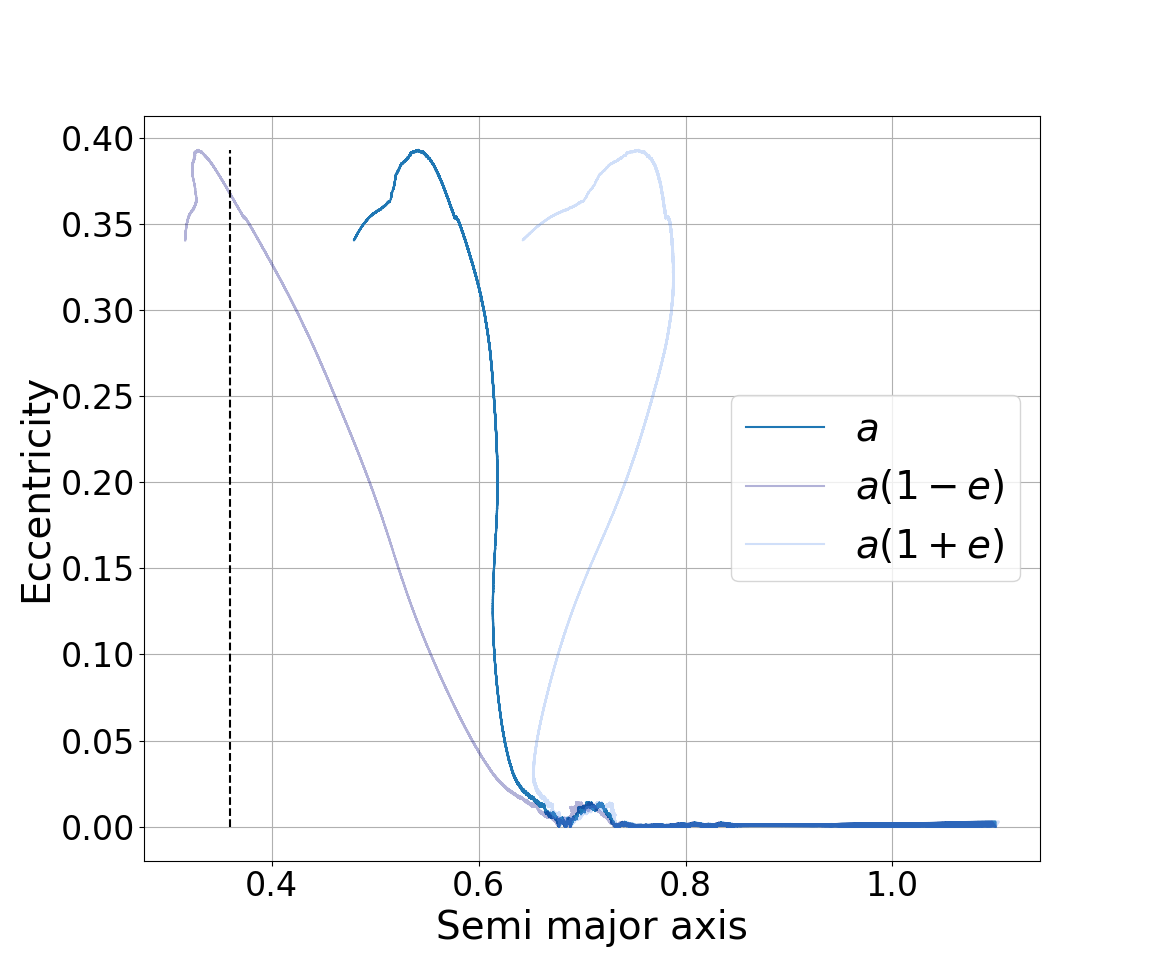}\label{fig:ref_2}
\caption{Left: semi-major axis and eccentricity of the planet as a function of time for the {\it 'Reference'} run. Right: eccentricity as a function of semi-major axis. For visual aid, the planet's pericentre, $a (1-e)$, and apocentre, $a (1+e)$, are plotted in transparency. The outer edge of the inner wave-killing zone is shown by a dark dashed line in both panels. 
}
\label{fig:reference}
\end{figure*}

\begin{figure*}
\centering
  \includegraphics[width=.48\linewidth]{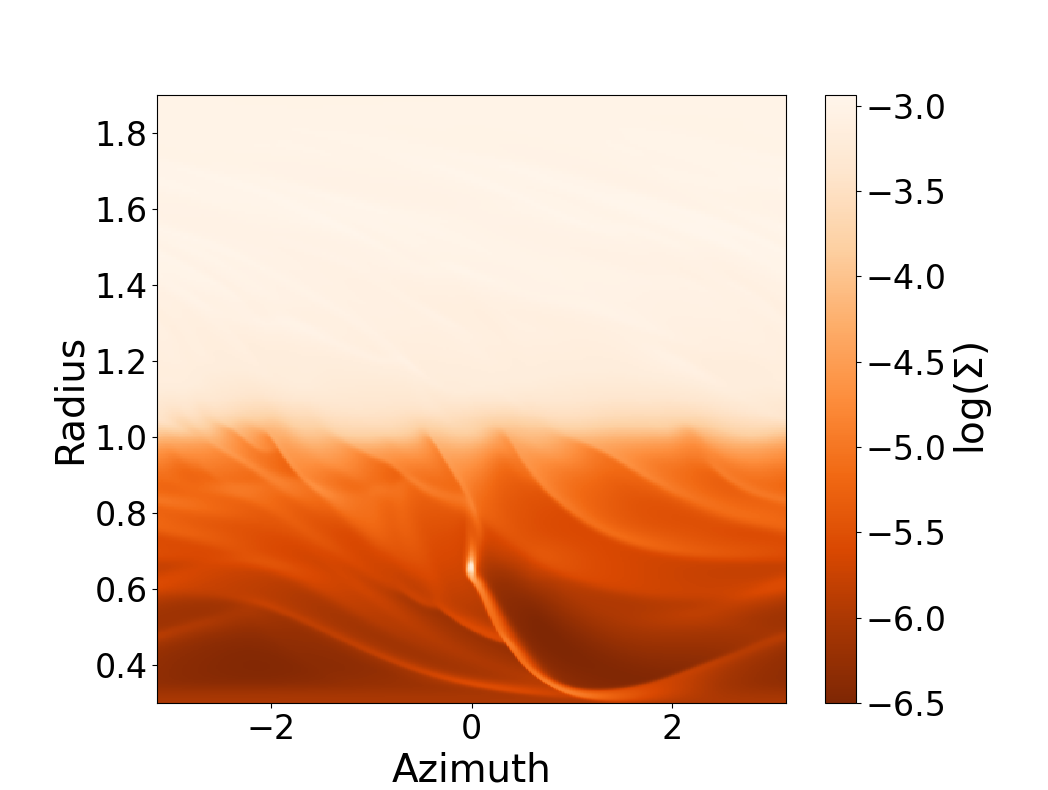}
  \includegraphics[width=.48\linewidth]{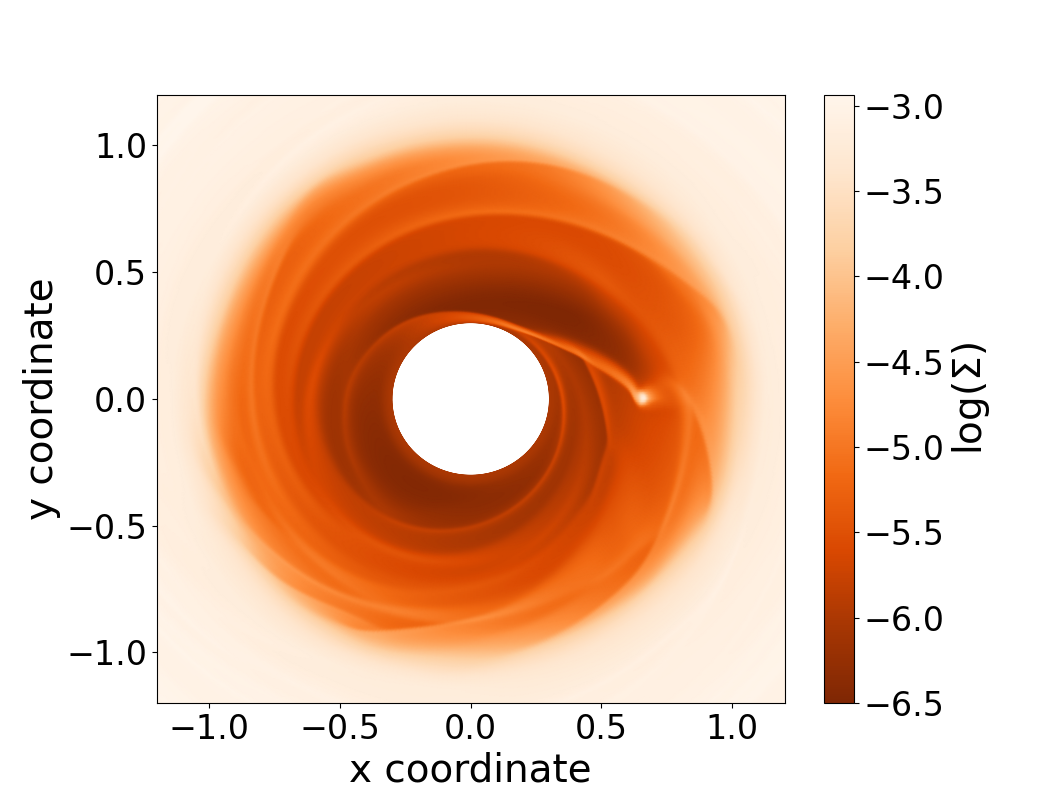}
\caption{Gas surface density in decimal log scale, shown in (left) polar and (right) cartesian coordinates after 18 000 orbits in the {\it 'Reference'} simulation.}
\label{fig:steady}
\end{figure*}

In the left panel of Figure \ref{fig:reference}, we show the time evolution of the planet's semi major axis and eccentricity in the {\it 'Reference'} run, while the right panel displays eccentricity versus semi-major axis. The evolution of the eccentricity and semi-major axis with time can be split in 4 steps: 

\begin{itemize}
\item During the first few thousand orbits, the inward migration is fast because of the contrast between the high density in the outer disc and the low density in the inner disc. The eccentricity essentially remains null, as expected for migrating Jupiter-mass planets. 
\item The inward migration then gets slower with time, which is expected as the planet enters the cavity where the disc density is lower.

\item After 5000 orbits, the planet experiences a slight increase in eccentricity. We attribute this to the crossing of the 2:1 resonance with a vortex formed by the Rossby wave instability \citep{Lovelace1999} at the outer edge of the cavity.

\item After $\sim$ 8000 orbits, the planet eccentricity increases monotonically with time. This occurs as the planet's semi-major axis becomes less than $\sim 0.68$ (in code units), which is close to $2^{-2/3}$: almost all the principal outer Lindblad resonances are now inside the cavity. When comparing both panels in Figure \ref{fig:reference}, we see that the slow increase with time of the planet eccentricity corresponds to an exponential increase of eccentricity with decreasing semi-major axis. The density of the disc after 18 000 orbits is shown in Figure \ref{fig:steady}.


\item When the eccentricity reaches $\sim$0.35, after 20000 orbits, the  planet enters the wave killing zone at the pericentre of its orbit, preventing to discuss the evolution on longer timescale. We discuss this effect in section \ref{ssec:bounchange}.
 \end{itemize}

In summary, the planet eccentricity in our reference run reaches values never obtained so far for Jupiter-mass planets solely through planet--disc interactions. The eccentricity reaches values that are even higher than the observed median eccentricity of warm Jupiters, which makes our mechanism a promising explanation for the origin of the eccentricity of warm giant planets, provided that central cavities with sizes of a few 0.1 AU to a few AU can be commonly generated  in protoplanetary discs (see Section \ref{sec:discussion}). We now turn to assessing the effects of the numerical and physical parameters on the evolution of the eccentricity.

\subsection{Changing the boundary conditions}
\label{ssec:bounchange}

\begin{figure*}
\centering
  \includegraphics[width=.49\linewidth]{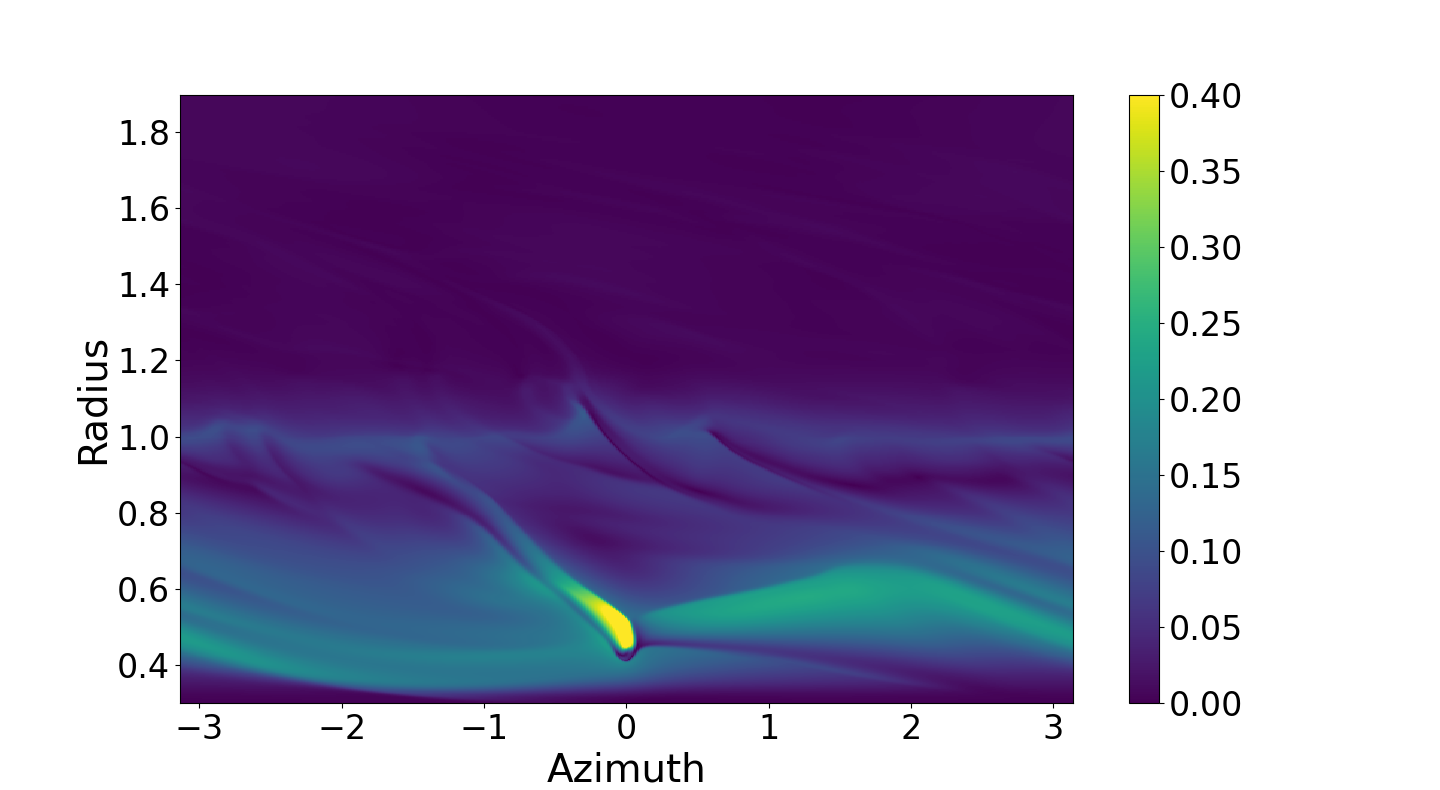}\label{fig:bound}
    \includegraphics[width=.49\linewidth]{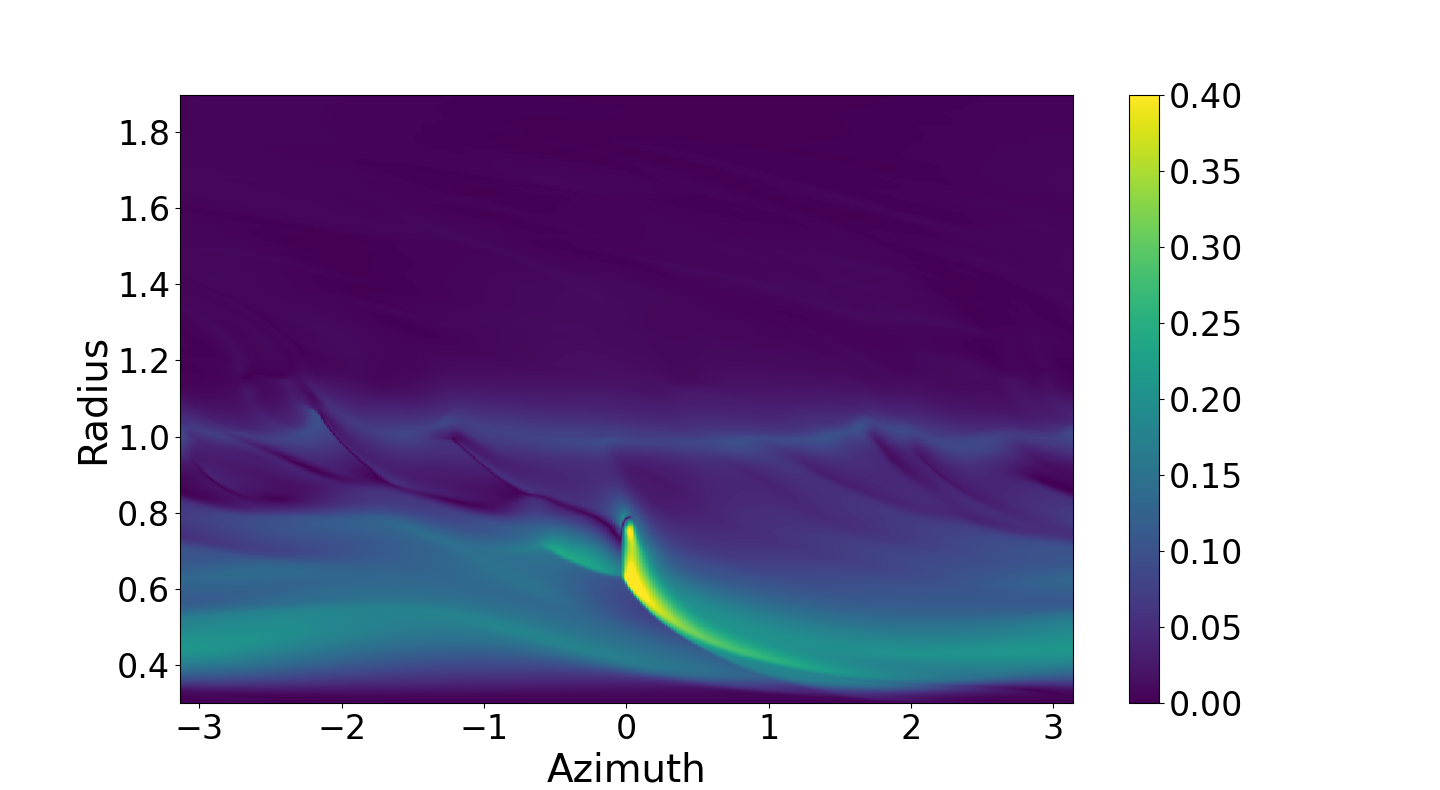}\label{fig:bound}
  \includegraphics[width=.49\linewidth]{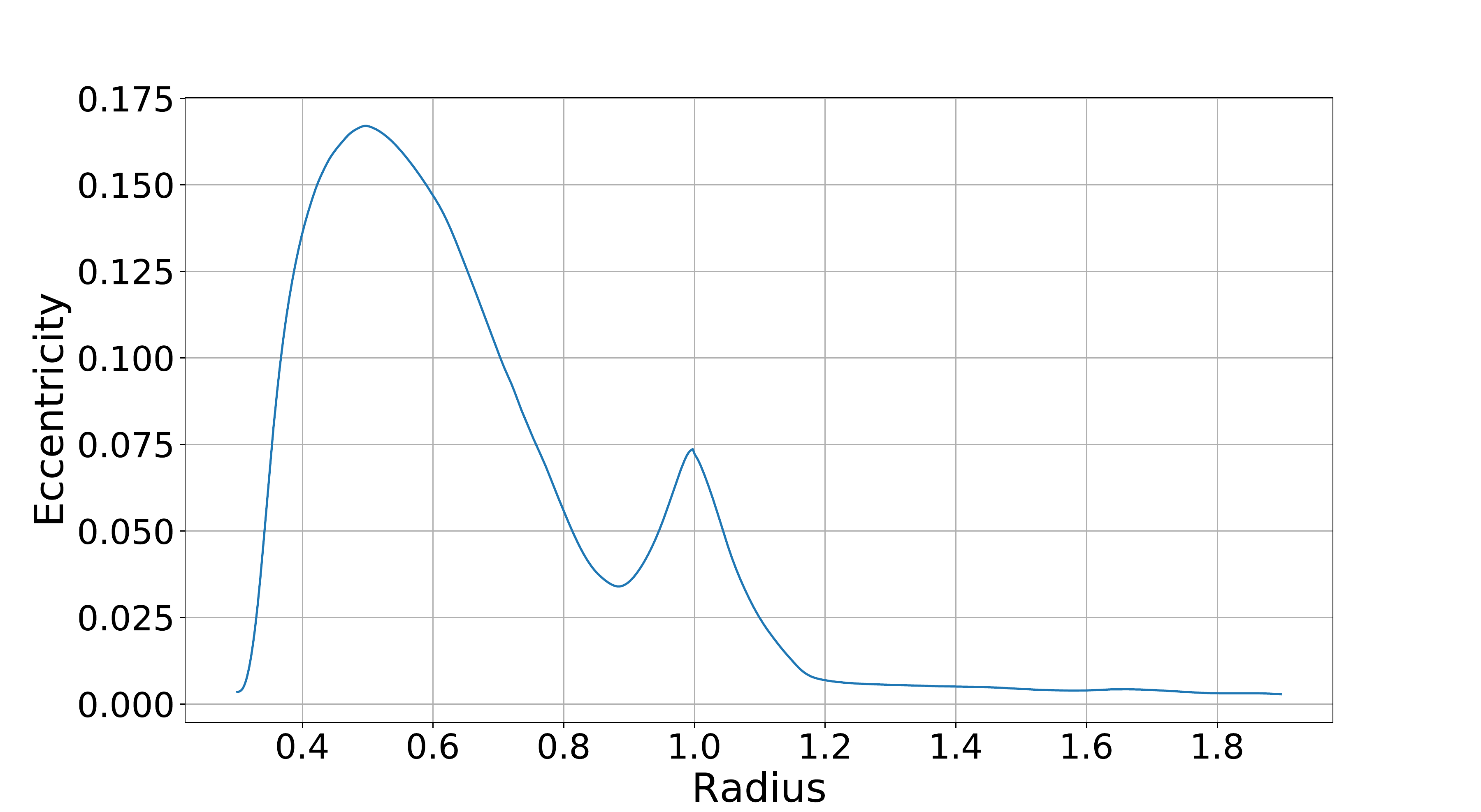}\label{fig:bound2}
\caption{Gas eccentricity in the {\it 'Reference'} simulation after 18000 orbits. Top: as a function of radius and azimuth at the pericentre (left) or apocentre (right) of the orbit. Bottom: azimuthally and temporally (2 orbits) averaged.}
\label{fig:ecc}
\end{figure*}

In the {\it 'Reference'} simulation, we cannot discuss on the final state of the simulation because of the interaction between the planet and the inner wave killing zone. In order to overcome this issue, we have run a very large set of simulations changing the size of the computational domain, the size of the wave killing zone and/or the 'ini' or 'axi' prescription in this zone (see \S \ref{ssec:boundary}). Unfortunately, a similar issue remains: either the planet enters the wave killing zone either the gas in the cavity gets highly eccentric and elliptic motions of the gas interact with the wave killing zone. Figure \ref{fig:ecc} shows the 2D and azimuthally averaged eccentricity of the gas in the {\it 'Reference'} simulation after 18000 orbits. It is clearly seen that the inner boundary condition imposes an almost null eccentricity in the wave killing zone.  That being said, all the simulations we have run while varying boundary conditions present an eccentricity growth up to $\gtrsim 0.2$ at least, and notably the {\it 'Reference'} simulation with "axi" boundary condition in the wave killing zone reaches as well an eccentricity superior to 0.3. 


In conclusion, the eccentricity growth we obtain is robust against a change in the prescription of the boundary condition. When the eccentricity of the planet gets too high, the boundary conditions prevent the eccentricity from getting larger. Therefore,  we cannot discuss on the maximal eccentricity that can be reached through this mechanism but  note that it could very well be much higher than 0.4. Dedicated simulations taking into account the interaction between the gas and the star would be needed, which is out of the scope of this paper.




\subsection{Convergence with resolution}
\label{sec:convergence}

\begin{figure*}
\centering
  \includegraphics[width=.49\linewidth]{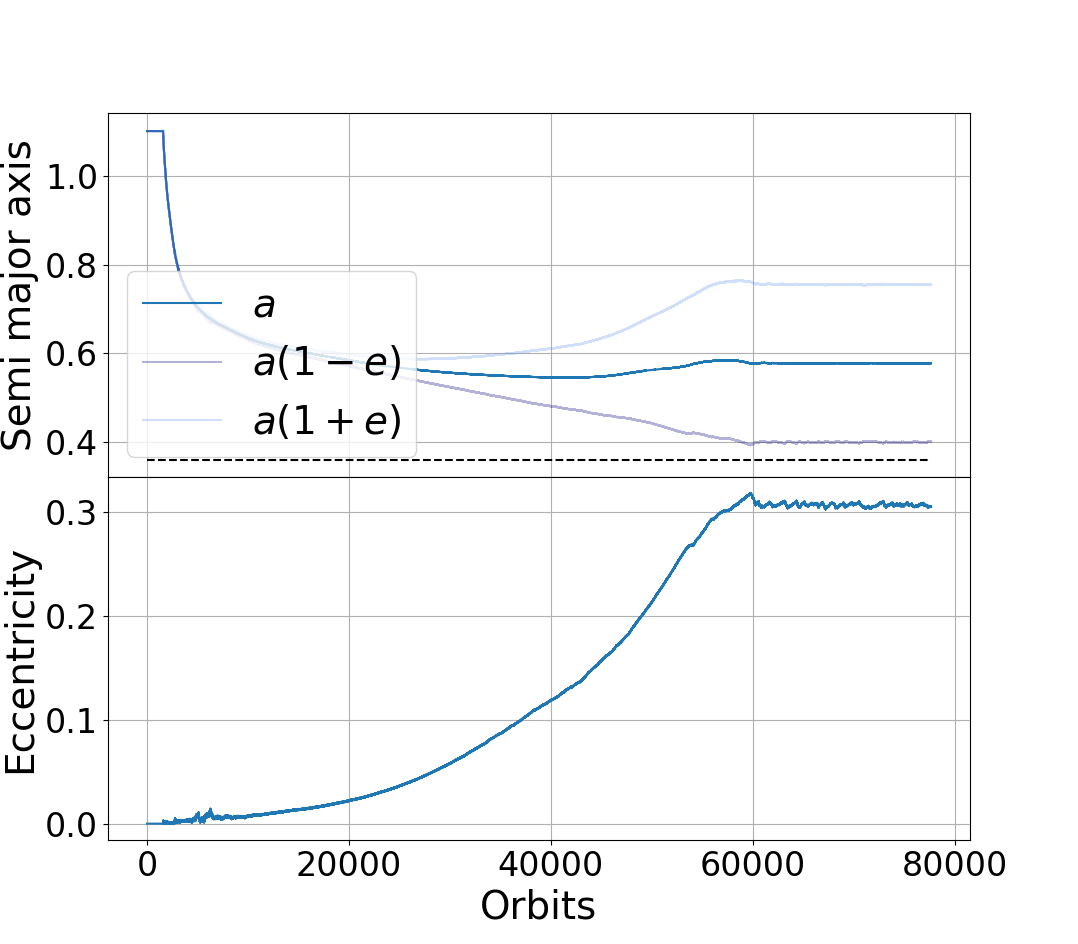}\label{fig:lr_2}
  \includegraphics[width=.49\linewidth]{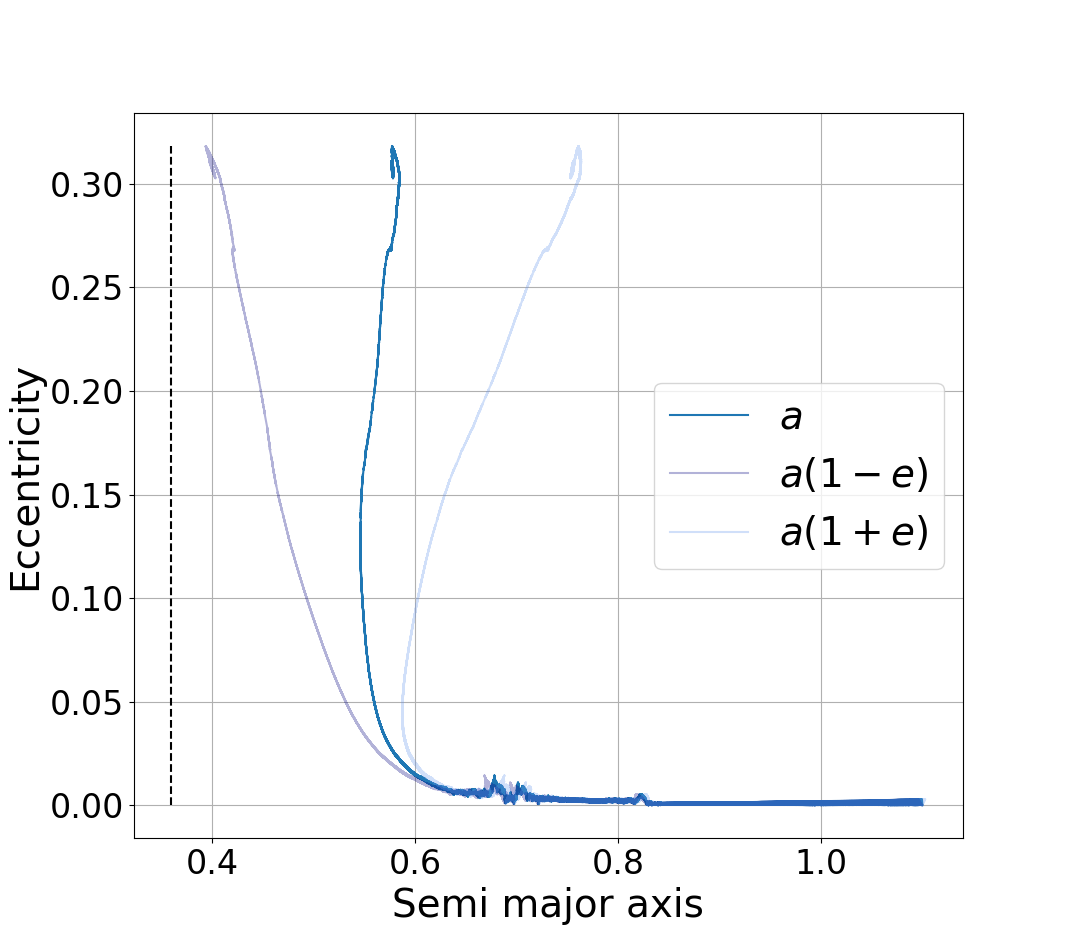}\label{fig:lr_1}
\caption{Same as Figure \ref{fig:reference} with a grid resolution of 200$\times$200.}
\label{fig:lr}
\end{figure*}

In order to assess the validity of the proposed mechanism, it is necessary to test its robustness against different resolutions. First, we have  run a low resolution simulation, with 200$\times$200 grid cells and all the other parameters similar to the {\it 'Reference'} run whose results are shown on Figure \ref{fig:lr}. The most striking differences are: (i) the eccentric instability is triggered at a different, smaller semi-major axis in the low resolution simulation and (ii) the characteristic time for eccentricity growth is about 3 times longer in the low res simulation. 

In order to understand these differences, we have run the {\it 'Reference'} and the low resolution simulations with both Fargo and FARGO3D. At low resolution, the two codes do find an eccentricity growth, but the initial position and time for the instability to occur are sensibly different (about 10 percent). However, at a resolution of 400$\times$400, the two codes are in almost perfect agreement on the evolution of the eccentricity as a function of time and position. We therefore conclude that, although the 200$\times$200 simulation captures the mechanism that leads to an eccentricity growth, it does not precisely evaluate the orbital evolution of the planet. Only qualitative conclusions can be drawn for such a resolution. 

\begin{figure}
\centering
  \includegraphics[width=\linewidth]{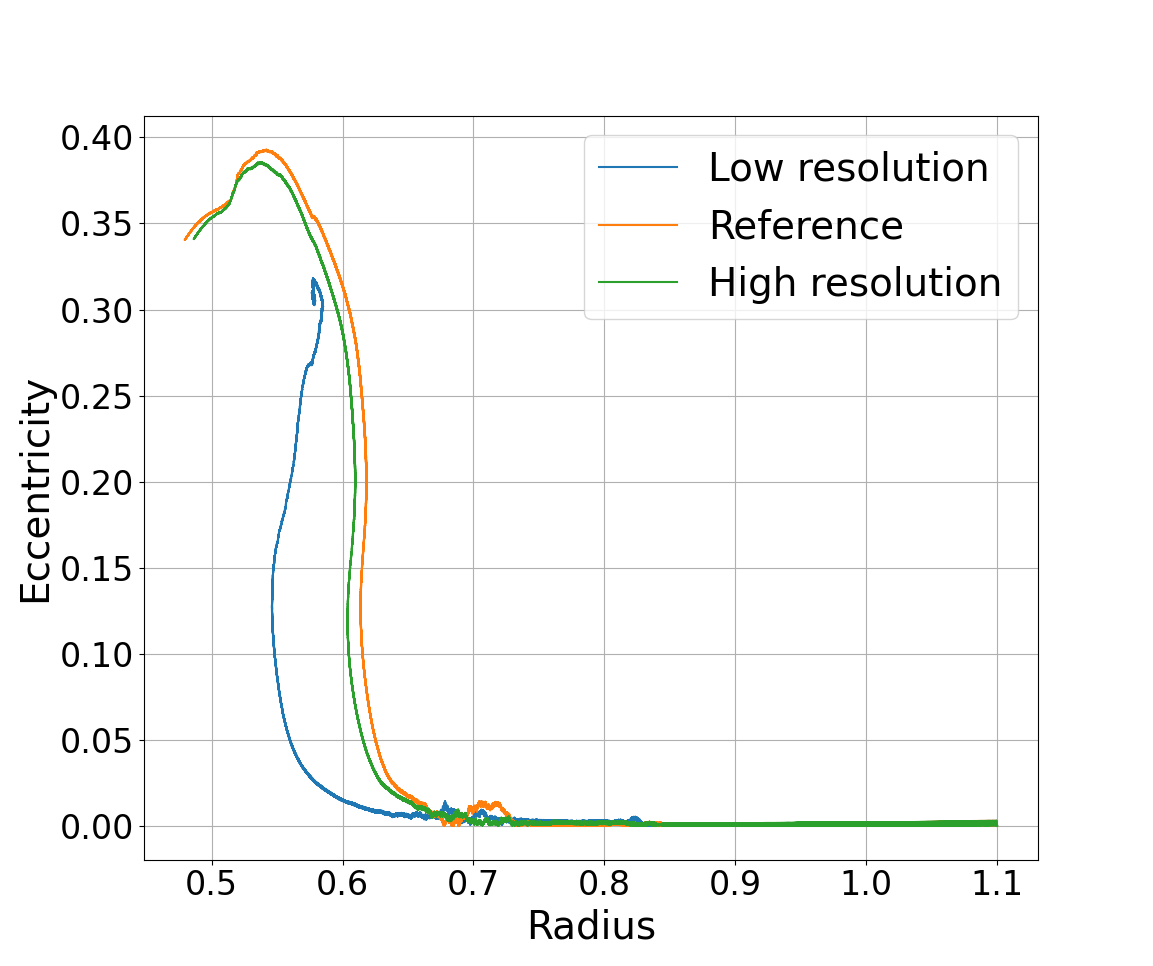}
\caption{Eccentricity as a function of semi-major axis for simulations with the parameters of the {\it 'Reference'} run except for the resolution: 200$\times$200 (low), 400$\times$400 (reference) and 600$\times$600 (high).}
\label{fig:resolution}
\end{figure}

We have finally run a higher resolution run (600$\times$600) in order to assess the convergence with resolution. Figure \ref{fig:resolution} shows the eccentricity of the planet as a function of semi-major axis for the three resolutions (200$\times$200, 400$\times$400 and 600$\times$600) with all the other parameters identical to the {\it 'Reference'} run. This figure shows that the difference between the simulations decreases with increasing resolution, confirming the numerical robustness of the eccentricity growth we obtained.

\subsection{Dependence on the physical parameters}
\label{ssec:physics}

In this section, we study the impact of the physical description of the disc on the eccentricity growth of the planet. As the {\it 'Reference'} run required about 40 000 hours of CPU time, it was not possible for us to explore the effects of the different parameters with such a numerical cost. We have therefore run our exploratory simulations at low resolution: 200$\times$200 cells in the radial and azimuthal directions, as in Figure \ref{fig:lr}.

First, we have varied the disc aspect ratio $h$ and the planet-to-star mass ratio $q$. The rationale behind this exploration is that the ability of a planet to carve an annular gap  through its shock wakes and build a massive circumplanetary disc depends on $q$, $h$ and on the disc's turbulent viscosity \citep[e.g.,][]{Crida2006}. A lower $h$ should lead to a stronger depletion of the planet's coorbital region, and should therefore favour growth of the planet's eccentricity. On the other hand, a higher $h$ would prevent the planet to carve a deep gap, which would tend to reduce the effect of the eccentricity pumping. We have carried out three simulations with varying $h$ (although keeping $h$ constant with radius). These three simulations have $h=0.03$, $h=0.07$, and $h=0.1$. 

For $h=0.1$, the inward migration of the planet stalls outside the cavity, and the planet therefore experiences no increase in eccentricity\footnote{The planet might still end up in a cavity without migrating: if 'inside-out' photoevaporation extends the cavity up to the planet's position. We have not tried such a setup in this work.}. However, we note that such a high aspect ratio implies either a very high temperature in the disc's inner parts (at a few AU) or that the cavity would be located at a few tens of AU in the disc, but in that case eccentricity growth might be too slow to develop (see Section \ref{sec:discussion}). In the simulation with $h=0.07$, the planet enters the cavity, and its semi-major axis stalls at $a \approx 0.62$, with a steady eccentricity of $0.08$. We have checked that the planet's orbit remains stable over 40 000 orbits, after what we stopped the simulation. We have used this case to test the effects of accretion on to the planet, with a simple prescription following \citet{Kley1999}, and the effect of a more physically motivated temperature profile in the vicinity of the planet (see e.g.,  \citet{Peplinski2008,Lin2012}). Overall, the results were qualitatively unchanged unless for an unrealistic accretion or temperature close to the planet. Finally, the $h=0.03$ simulation is broadly similar to the $h=0.05$ simulation, although the eccentricity growth occurs at earlier times.

For completeness, we have also run three additional simulations with $h=0.05$ but where the planet mass was changed to have the same $q/h^3$ ratio as in the simulations presented in the previous paragraph (where $h$ was varied at fixed $q$). The reason for this exploration is that $q/h^3$ is {\it a priori} one of the important (dimensionless) parameters that control the width and depth of planet gaps. We thus examine here the effect of varying $q$ at fixed $q/h^3$ on our mechanism for eccentricity growth. The three values of $q$ are $ 1.2\times 10^{-4}$, $3.6\times 10^{-4}$ and $4.6\times 10^{-3}$. For the two lower values of $q$, no eccentricity growth is observed. Similarly to the case with $q=10^{-3}$ and $h=0.1$ in the previous paragraph, the planet with $q =1.2\times 10^{-4}$ in the $h=0.05$ disc stalls its migration before entering the cavity. Interestingly, the run with $q =3.6\times 10^{-4}$ and $h=0.05$ shows no eccentricity increase while that with $q=10^{-3}$ and $h=0.07$ does. Lastly, the run with $q =4.6\times 10^{-3}$ and $h=0.05$ does experience eccentricity growth comparable to the $q =10^{-3}$ and $h=0.03$ case. These results highlight that the $q/h^3$ ratio is not the only quantity to assess the possibility of eccentricity growth in a cavity, as we further discuss in Section \ref{ssec:terminal}.

Interestingly, we obtained an eccentricity growth up to $\sim 0.25$ for a planet of mass $M_\mathrm{J}/2$ in low and mid resolution for a disc aspect ratio of $h=0.03$. This points towards the fact the proposed mechanism can be extended to lower masses, although we have not explored whether this is robust against a change in the other physical parameters.

Considering the density contrast between the cavity and the outer disc, we have run four additional  simulations by changing the value of $\Sigma_c$ and $\Sigma_0$  (we recall that $\Sigma_c = 10^{-6}$ and $\Sigma_0 = 10^{-3}$ in the {\it 'Reference'} run). (i) When $\Sigma_c = 2 \times 10^{-6}$, the planet experiences a very similar evolution than in the low resolution case. The typical time and position for eccentricity evolution differ by only a few percent, and the final position is the same. (ii) When $\Sigma_c = 5 \times 10^{-6}$, the planet's eccentricity increases and settles at 0.02 while the semi major axis keeps decreasing with time. (iii) When $\Sigma_0 = 5 \times 10^{-4}$, the time for the migration and eccentricity evolution is 2 times longer, as expected, but the evolution of the eccentricity is sensibly similar to the low resolution case. (iv) Finally, when $\Sigma_0 = 2 \times 10^{-4}$, the results are comparable to the  $\Sigma_c = 5 \times 10^{-6}$ case: the planet's eccentricity increases very slightly and settles (up to 0.015 in this case) but does not reach high values. It is therefore quite clear than when the density contrast is not high enough, the threshold for the eccentric instability is not reached and the planet retains a low level of eccentricity. We discuss this further in the next section.

We have, however, not attempted to increase the initial density profile of the outer disc. Increasing the gas density outside the cavity by an order of magnitude would cause the outer disc to become gravitationally unstable, which would have required to include gas self-gravity.

Regarding viscosity, we have run a simulation where the viscosity was lowered by an order of magnitude throughout the disc. Its outcome turns out to be quite similar to that of the {\it 'Reference'} run. We have also run a simulation where $\Sigma_0 \nu_0 = \Sigma_1 \nu_1$ by decreasing $\nu_0$ by a factor of 10, in order to have an equal viscous mass flux through the boundaries. Once again, we recovered an eccentricity increase up to $\sim 0.3$ values, as in the low resolution simulation. This confirms that eccentricity growth via migration in a disc cavity is not simply due to a specific choice of viscosity. We have also varied the size of the viscosity transition $\Delta r$ as well as the viscosity prescription (see Eq. \eqref{eq:visc}). The main effects were to alter the position of the initial eccentricity growth, as it changes the density profile of the inner disc (inside of $r \sim 1.1$) and its magnitude, but eccentricity growth was almost ubiquitous.

Globally, the results of this section show that the mechanism we obtain for eccentricity growth does not rely on one  particular physical description of the disc, but is robust against a change in the parameters. Additionally, this mechanism can be extended to lower mass planets provided the disc's aspect ratio is low enough. The proposed mechanism is therefore a very promising explanation for the origin of the eccentricity of planets ranging from Saturn mass to a few Jupiter mass.

\section{Discussion}
\label{sec:discussion}

\subsection{Comparison with analytical estimate}
\label{ssec:analytics}

In this section, we compare the results we obtained with simple analytical estimates. Because the assumptions used to derive the analytical torque are not similar to our setup (notably the consideration of a non-zero viscosity), the following estimations are quite crude, but show a good comparison with numerical results. We do not discuss either on the potential saturation of the corotation resonances, that would increase the speed of eccentricity growth. 

Following \citet{Masset2008}, energy and angular momentum conservations lead to a simple equation for the evolution of the eccentricity of a planet, given a momentum flux (or torque) $T$ (their equation (5.9)): 
\begin{equation}
    \dfrac{M_\mathrm{p}}{2}\dfrac{\Omega_\mathrm{p} a^2}{\sqrt{1-e^2}}  \dfrac{\mathrm{d} e^2}{\mathrm{d} t} = \left(1 - \dfrac{\Omega_\mathrm{d}}{\Omega_\mathrm{p}}\sqrt{1-e^2} \right) T
    \label{eq:ecc_growth}
\end{equation}
where $\Omega_\mathrm{d}$ is the pattern frequency of the perturbing potential (see \citet{Masset2008}) and $\Omega_\mathrm{p}$ the planetary frequency. 
For low eccentricity, this leads to a characteristic time $\tau$ for eccentricity growth (note that the torque we are interested in scales as the square of the eccentricity, \citet{GoldreichTremaine}): 

\begin{equation}
    \tau \sim \dfrac{M_\mathrm{p} \Omega_\mathrm{p} a^2}{\left(1 - \dfrac{\Omega_\mathrm{d}}{\Omega_\mathrm{p}} \right) (T/e^2)}
    \label{eq:time_growth}
\end{equation}

As detailed in \citet{GoldreichTremaine,Goldreich1980}  (see also our introduction or \citet{Goldreich2003}), the leading resonances that excite or damp eccentricity are the first order resonances, for which 

\begin{equation}
    \Omega_\mathrm{d} = \Omega_\mathrm{p} \left(1 \pm \dfrac{1}{m} \right)
\end{equation}
where $m$ is the index of the decomposition of the planet potential into an azimuthal Fourier series. A
torque can only be exchanged between the planet and the disc at Lindblad or corotation resonances, where:

\begin{equation}
    \Omega(r) \left(1 + \dfrac{\epsilon}{m} \right)= \Omega_\mathrm{d} 
\end{equation}
with $\epsilon = 1$ for inner Lindblad resonances (ILR), $0$ for corotation resonances (CR) and $\epsilon = -1$ for outer Lindblad resonances (OLR). 
Because of our density profile that sharply increases with radius through the cavity, the dominant resonances during the eccentricity growth in our simulations are the CR and OLR of the "slow term" (see Figure 22 of \citealt{Masset2008}), for which the positions $r_\mathrm{OLR}$ and $r_\mathrm{CR}$ are:
\begin{gather}
    r_\mathrm{OLR} = r_\mathrm{p}\left(\dfrac{m+1}{
m-1}\right)^{2/3} \text{,} \\
    r_\mathrm{CR} = r_\mathrm{p}\left(\dfrac{m}{
m-1}\right)^{2/3}
\end{gather}
This is in sharp contrast with usual consideration for Jupiter mass planets, where the ILR of the slow term and the OLR of the fast term have a dominant effect on the damping of eccentricity. On the other hand, this is the same picture than \citet{Papaloizou2001} where such resonances fall into the planetary gap.

\begin{figure}
\centering
  \includegraphics[width=\linewidth]{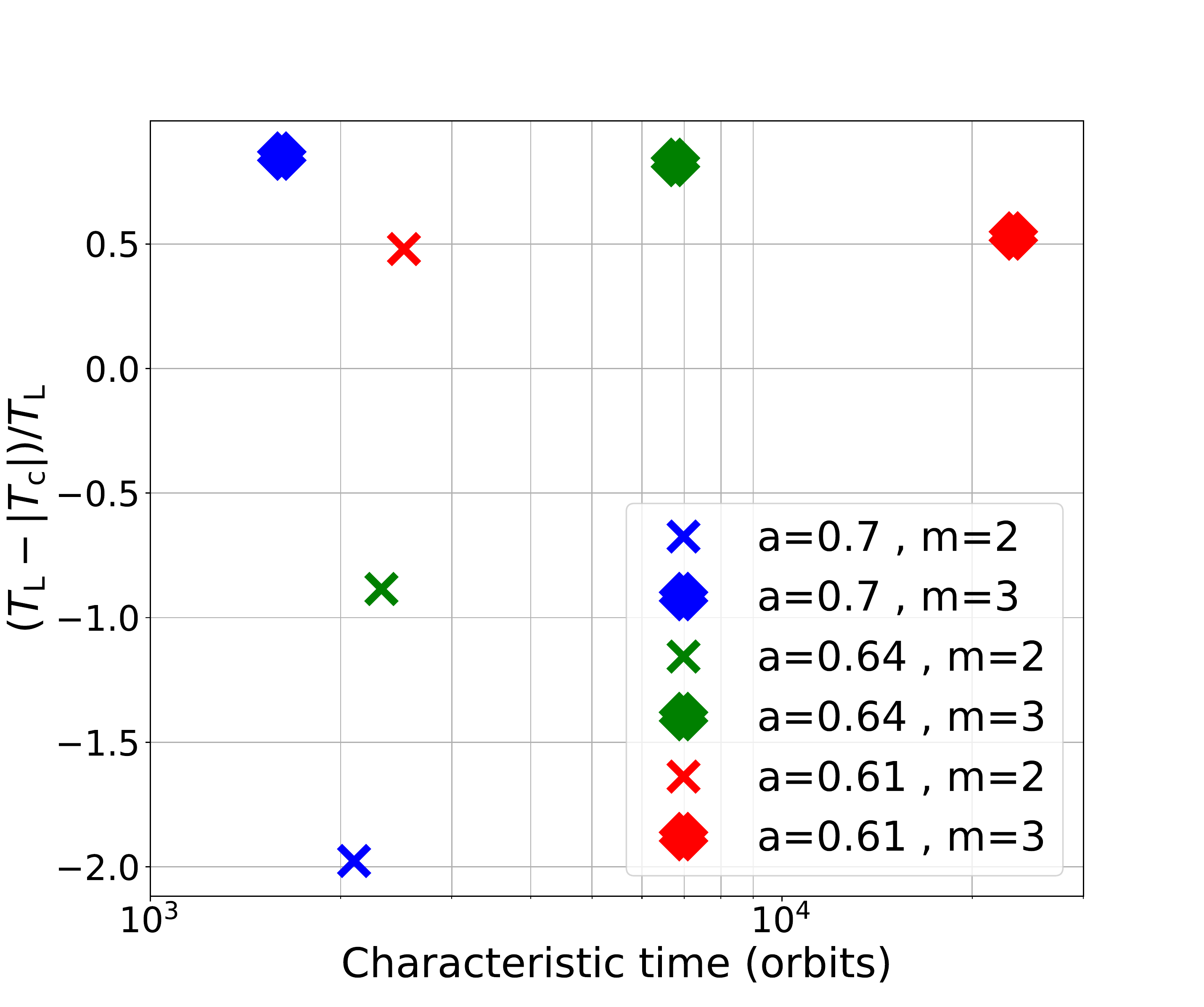}
\caption{Difference in percentage between the Linbald ($T_\mathrm{L}$) and corotation ($T_\mathrm{C}$) torque, as a function of the characteristic time for eccentricity growth considering only the Linbald torque (see Eq.\eqref{eq:time_growth}, inversely proportional to the torque). Data are taken from the high resolution simulation, after 5370, 10 0000 and 15 0000 orbits, corresponding to $a=0.7$, $a=0.64$ and $a=0.61$ respectively. The shape of the symbol indicate the order of the Fourier series.}
\label{fig:ecc_growth}
\end{figure}

The OLR of the slow term is known to lead to an increase of the eccentricity, however the effect of the CR depends on the gradient of $\Sigma/\Omega$ (\citealt{Goldreich1980}, equation 14). In the transition region of our simulations (where this CR lies), this gradient is positive: the CR leads to a damping of eccentricity, in contrast with the OLR. In order to determine which of these resonances is dominant, we therefore calculated numerically the torque exerted at the CR and OLR of the slow term during the evolution of the planet in the high resolution simulation, for $m=2$ and $m=3$ ( assuming that the density profile is correctly represented by the azimuthally averaged density). These are the dominant components, as when $m$ increases, $r_\mathrm{OLR}$ and $r_\mathrm{CR}$ decrease and move to regions with much smaller densities. The results are shown in Figure \ref{fig:ecc_growth}.

At early time, (blue color in Figure \ref{fig:ecc_growth}) although the $m=3$ component would lead to an increase in the eccentricity, the corotation torque of the $m=2$ component is too strong and prevents an eccentricity growth. At later times, the $m=3$ component  decreases in strength because it lies in the region where the density decreases sharply with decreasing radius (green color). However, the density in the vicinity of the Lindblad, $m=2$ resonance becomes much higher than that of the corotation resonance. This leads to a reversal of the situation at $a=0.61$ (red cross): the cumulative effect corresponds to a positive torque, hence an increase in the eccentricity (see Eq.\eqref{eq:ecc_growth}). The characteristic time for this increase is of the order of a few thousands orbits, which is very comparable to the outcome of our simulations (Figure \ref{fig:reference}). On the other hand, the  $m=3$ component decreases in strength down to a tenth of  the $m=2$ component when $a=0.61$.  Therefore, this points towards the fact that, as in \citet{Papaloizou2001} but unlike \citet{Dangelo2006}, the 1:3 resonance has the major effect in the eccentricity growth of our simulations and is able to excite eccentricity on a characteristic time of a few thousand years. We attribute this effect to our specific choice of density in the cavity, that leads to a sharp density gradient hence enhances the effect of resonances that are located far from the planet.  

\subsection{Carving a cavity and disc dispersion}
Throughout this paper, we have assumed that a cavity already exists in the disc and that its shape is unchanged by any effects other than planetary migration. A possible origin for disc cavities is FUV or X-ray photo-evaporation from the star \citep[see ][]{Alexander2006,Gorti2009,Owen2011,Ercolano2018,Picogna2019,Owen2019}.  In this case, a gap is created at a distance of up to a few AU from the star where X-ray heating is most efficient, which leads to a decoupling between the outer disc and the inner disc. The inner disc then empties out by viscously accreting onto the star (and eventually through winds), while the outer disc is rather depleted by further inside-out photoevaporation. Once a gap is created, the dispersion of the inner disc is quick (of the order of a few $10^4$ years, \citet{Ercolano2017-r}) while the complete dispersion of the outer disc is less constrained. 

The cavity could also be carved by the presence of magnetically driven winds \citep{Bai2016,Wang2017} which might enhance angular momentum loss in the inner parts of the disc. Although not as efficient as photo-evaporation in creating a cavity, disc winds do affect the density structure close to the star. A combination of photo-evaporation and magnetic winds is therefore a likely scenario to carve a cavity fast enough for planets to migrate inside, but the typical size of such cavities is not easily determined. 

Alternatively, the cavity could be depleted by another giant planet, as argued in \citet{Papaloizou2001}. If a second giant planet reaches the cavity and experiences eccentricity growth, gravitational interactions with the inner planet that built the cavity in the first place could occur, which could destabilise the system. This leads us to suggest that eccentricity growth in a cavity is a more likely viable scenario when the cavity is not built up by another companion.

Regarding the typical timescale for eccentricity growth, our simulations show that it takes about $10^4$ orbital periods to reach eccentricities superior to $0.05$ for our {\it 'Reference'} run. Therefore, if the cavity is 1 AU wide, the timescale for eccentricity growth is short and the mechanism is likely to occur.  If it is 5 AU wide, eccentricity growth will take about $10^5$ years and the outer disc might start to dissipate during the process, which would alter the final state of eccentricity growth. Further than 20 AU, the process would require at least $10^6$ years to yield a high eccentricity. The outer disc would probably have been emptied during this time, impeding the mechanism. Therefore, we do not expect highly eccentric giant planets through this mechanism further than $10-20$ AU around a Sun-mass star. The secular interaction mechanism of \citet{Ragusa2018}, which requires about $10
^5$ orbits in a low density disc, would also be too slow to develop further than 20 AU.  This prediction might prove to be an observational test of our mechanism in the future.

Interestingly, there seems to be an excess of  planets in the Neptune to Saturn mass range at large orbital distances, as evidenced by microlensing  observations \citep{Suzuki2016,Suzuki2018}. If cavity carving is a common process in protoplanetary discs, migration in a cavity could be a way to restrain the runaway gas accretion \citep{Pollack1996} of Neptune mass planets: the low gas  density in the cavity would not allow them to reach the mass of Saturn. In this scenario, the detected desert of  sub-Saturn mass planets at short orbital distances ($\lesssim $10 AU, see Figure 1) is not a confirmation of runaway gas accretion, but an evidence that planets in the Neptune to Saturn mass range stall their migration at a few AU from their star.  Such a conclusion is obviously very speculative at this point, but it provides an additional strength to the possibility that migration in disc cavities is a common process for planets across the galaxy.

\subsection{Maximum eccentricity and planetary migration}
\label{ssec:terminal}

We have seen through the paper that the migration in a cavity can increase the eccentricity of warm Jupiters up to $0.4$. However, the interaction with the inner wave killing zone limited the eccentricity growth. It is therefore possible that this mecanism leads to much higher  eccentricities, and only dedicated simulation with the interaction between the eccentric cavity and the star could help answering this question. 

It is interesting in that regard to write the eccentricity evolution for the planet at the next order in eccentricity than Eq.\eqref{eq:ecc_growth} (see \citet{Goldreich1980}, their equation 27, assuming a Keplerian disk):

\begin{equation}
    \dfrac{\mathrm{d}e}{\mathrm{d} t} \propto - T \left[ (\Omega_\mathrm{d} - \Omega_\mathrm{p}) + e^2 \Omega_\mathrm{p}  \right ] 
\end{equation}
where, again, $T$ is the torque. 
The second term of the right hand side can overcome the first term for high eccentricity and is of opposite sign: a positive torque would then lead to a damping of eccentricity. For the slow term (see section \ref{ssec:analytics}), this requires $e = \sqrt{1/m}$, which is $e =  0.7$ for $m=2$. If the $m=2$ component is dominant at all times, this effect will not prevent eccentricity growth to much higher values than 0.3 (although the linear approximation breaks before $e=0.7$).

Additionally, the maximum eccentricity that a warm Jupiter can reach in our simulations is influenced by the density ratio between the cavity and the outer disc. This ratio depends on the efficiency of the physical mecanism that carved the cavity to maintain a low density in the cavity while preventing the outer gas to viscously refill that cavity. Only when this ratio is higher than $500$ did we obtain eccentricities larger than $0.15$. As the median eccentricity of warm Jupiters with planet-to-star mass ratio q satisfying $5 \times 10^{-4} < q < 5 \times 10^{-3}$ is 0.2, this might provide hints about the level of emptiness of cavities in protoplanetary discs, and therefore on the physics of protoplanetary discs.


As seen in Section \ref{ssec:physics}, our mechanism could be applied to explain the eccentricity of Saturn-mass planets if they evolved in a low aspect ratio region of the disc. Regarding lower mass planets, we cannot predict the results as we ran low resolution simulations where the horseshoe region is not properly resolved, although the mechanism can still occur conceptually. Overall, we think that our mechanism can be a simple and robust phenomenon to understand the origin of the eccentricity of planets with: (i) masses between the mass of Saturn and a few Jupiter mass ($ 3\times 10^{-4} \lesssim q \lesssim 5 \times 10^{-3}$) and (ii) intermediate value for the eccentricity of warm Jupiters ($0.1 \lesssim e \lesssim 0.5$). However, care must be taken if one tries to explain the partition of eccentricities of lower mass planets through cavity migration, and extreme values for the eccentricity probably need to be explained through other mechanisms.

One of the interesting implications of the migration of a giant planet inside a cavity is that the final position of the planet depends both on the time the disc takes to carve and expand a cavity and the speed of planetary migration. We might therefore expect two different planet populations: when the formation and migration of the planet is faster than the time to carve a cavity, migrating warm Jupiters will end up very close to their star because of the standard type II migration paradigm \citep[e.g.][]{GoldreichTremaine,Lin1986,Robert2018,Baruteau2014}. This would be in line with the population of observed warm Neptunes and hot Jupiters with low projected orbital obliquities. On the other hand, if the time for the planet to migrate is longer than the time for the cavity to be carved, inward migration should stall at the location where the planet enters the cavity. Although other phenomena might stop the inward migration of planets, the persistence of a cavity is an appealing solution to explain the different partitions of planets with radius \citep[see discussions in][]{Alexander2012,Wise2018}. Therefore, if migration within a cavity is a common process in protoplanetary discs across the galaxy, the final position of planets could provide hints about the typical timescales for planet formation, migration and cavity opening in protoplanetary discs.

 

Finally, it is worth noting that in many planet population synthesis models \citep[see the review by][]{Benz2014}, the timescales for type I and II migration are often too short, and need to be increased somehow. For type I migration, the inclusion of thermal effects \citep{Masset2010,Paardekooper2011} leads to an increase in the migration time, but no equivalent has been obtained for type II migration. It seems quite clear that in any case, the presence of a cavity can only lead to slowing down or stopping the inward migration when/if the planet enters the cavity. This could provide a natural, physical explanation to increase further the timescale for type I and II migration.

\subsection{Hot Jupiters and Magnetospheric Gap}
\label{ssec:hotJ}
The magnetic interaction of a star with its disc leads to the formation of a magnetospheric cavity 
that can be of the order of a few stellar radii, typically 0.05 to 0.1 AU (as evidenced in the series of papers by Ghosh and Lamb: \citet{Gosha,Goshb,Goshc}, reviewed in, e.g., \citealp{Bouvier2007,Romanova2015}, and see also observations by \citet{Menard2003,Donati2008}).

The migration of low-mass planets inside the magnetospheric cavity is hard to predict because of the unknown or unconstrained strengths of the disc's magnetic field and magnetic resistivity \citep{Guilet2013}. However, gap-opening Jupiter-like planets, for which the corotation torque should be very small, are expected to enter 
deep inside the cavity.

Although most hot Jupiters have a circular orbit, 
some of them retain an eccentricity $\sim 0.1$ \citep{Kane2012,Shabram2016}. The most likely explanation is that these planets have a high tidal quality factor, leading to a long time for circularization and we observe them in the phase of circularisation \citep{Matsumura2008}. The origin of their initial eccentricity, however, is unknown. Interactions with an external perturber is, as usual, the favoured consequential explanation, but \citet{Matsumura2008} note that such perturbers should have been observed in the majority of the systems, although they are not. Notably, \citet{Knutson2014} performed a search for companions to hot Jupiters, and noted that "[they] find no statistically significant difference between the frequency of companions to transiting planets with misaligned or eccentric orbits and those with well-aligned, circular orbits", making planet--planet scattering a less favoured explanation for hot Jupiters' eccentricities. The same conclusions holds for an eventual interaction with a companion star \citep{Ngo2015}.

Therefore, eccentricity growth via migration in a disc cavity could provide a natural explanation so as to make these planets originally eccentric whereas tidal interactions would circularize orbits on a longer timescale once the gas disc is dispersed, as proposed by \citet{Rice2008}, \citet{Teyssandier2016} or \citet{Ragusa2018}). Our work provides additional strength to this scenario where hot Jupiters acquire their eccentricity after migrating in the magnetospheric cavity through planet-disc interactions.

\subsection{Inclination}

Among the possible extensions to this work, going to three-dimensional simulations seems very relevant, not only to check the level of eccentricity growth in such simulations, but also to investigate how the planet's inclination would evolve in the cavity. It is indeed possible that such a mechanism to excite eccentricity might have an effect on the inclination of planets. Therefore, although our values for the eccentricity compare well with the observed distribution of eccentricities of warm Jupiters, having access to a statistical repartition of inclination could be another way to confirm the viability of our proposed mechanism, or else assess its limitations. The numerical cost of such a simulation might however be prohibitive thus far.

\section{Conclusions}
In this paper, we have proposed a simple mechanism to pump up the eccentricity of warm Jupiters up to their mean observed eccentricity: migration inside a cavity.
We relied on the study of \citet{Papaloizou2001} that showed that planets of masses $\gtrsim 10 M_\mathrm{J}$ carve such a wide gap around their orbit that the first order corotation resonances have almost a null contribution on the orbital evolution of the planet, whereas the eccentric resonances get dominant. We therefore adapted this mechanism to warm Jupiters of lower masses ($0.5 - 5 M_\mathrm{J}$), in a disc that already exhibits a cavity. This cavity acts at reducing the effects of the principal Lindblad and first-order corotation resonances and prevents the gas close to the planet to force the planet on a circular orbit because of its low density.

We modelled the cavity as a jump in viscosity between the inner and outer parts of the disc, and allowed the planet to migrate inward in this cavity. We used a simplified treatment of the physics (2D disc model, locally isothermal equation of state, no accretion on the planet, no self gravity of the gas) and a middle or low resolution in order to run simulations for $\gtrsim \text{few } 10^4$ orbits, the minimum time for eccentricity growth to occur. Assessing the viability of the mechanism in a general case would have been prohibitively long for our exploration of the parameter space.

We obtained values for the eccentricities of warm Jupiters up to $e \sim 0.4$. Such a high eccentricity had never been obtained solely through planet--disc interactions, and we provide a robust mechanism to explain how warm Jupiters reach such level of eccentricity without invoking (poorly constrained) planet--planet scattering. We have assessed the effect of the  numerical parameters of the simulations, and confirmed their convergence with resolution.

Regarding the physical description of the simulations, we have shown that our proposed mechanism can extend to Saturn mass planets provided that the aspect ratio of the disc is decreased compared to our {\it 'Reference'} run ($h = 0.05$) and that, on the other hand, we do not expect the mechanism to work when the aspect ratio is larger than $h \sim 0.08$ for a Jupiter-mass planet. This is due to the fact that the planet stops migrating before entering far enough in the cavity for the eccentric resonances to get dominant. We also find that the viscosity does not affect much our results compared to the impact of the density and aspect ratio of the gas.

Overall, we propose a simple, robust mechanism to pump up the eccentricity of warm Jupiters, and have assessed the effect of changing our physical description and numerical parameters. This mechanim extends to a range of masses covering Saturn to super-Jupiter planets, and implies that cavity carving could be a very efficient mechanism in protoplanetary disc. It is unlikely that this mechanism alone explains the distribution of radii and eccentricities of all warm Jupiters, but we expect it to have a major effect on their statistical repartition.
\section*{Acknowledgments}
We thank Cl\'ement Ranc for stimulating discussions and an anonymous referee their valuable comments. This project received funding from the European Research
Council (ERC) under the H2020 research \& innovation programme (grant agreement \#740651 NewWorlds). Most of the numerical simulations were performed on the CALMIP Supercomputing Centre of the University of Toulouse. This work was granted access to the HPC resources of IDRIS under the allocation 2019-A0070410970 made by GENCI.

\section*{Data Availability}
All of the simulation data and python codes for plotting are available upon request to the corresponding author.

\bibliographystyle{mnras}
\bibliography{biblio_eccentric}

\bsp	
\label{lastpage}
\end{document}